# Two-photon interference from remote deterministic quantum dot microlenses

A. Thoma,[1] P. Schnauber,[1] J. Böhm,[1] M. Gschrey,[1] J.-H. Schulze,[1]
A. Strittmatter,[1,a)] S. Rodt,[1] T. Heindel,[1,b)] and S. Reitzenstein[1]

[1]*Institut für Festkörperphysik, Technische Universität Berlin, Berlin, 10623, Germany*

We report on two-photon interference (TPI) experiments using remote deterministic single-photon sources. Employing 3D in-situ electron-beam lithography, we fabricate quantum-light sources at specific target wavelengths by integrating pre-selected semiconductor quantum dots within monolithic microlenses. The individual single-photon sources show TPI visibilities of 49% and 22%, respectively, under pulsed p-shell excitation at 80 MHz. For the mutual TPI of the remote sources, we observe an uncorrected visibility of 29%, in quantitative agreement with the pure dephasing of the individual sources. Due to its efficient photon extraction within a broad spectral range (> 20 nm), our microlens-based approach is predestinated for future entanglement swapping experiments utilizing entangled photon pairs emitted by distant biexciton-exciton radiative cascades.

---

a) Present address: Abteilung für Halbleiterepitaxie, Otto-von-Guericke Universität, 39106 Magdeburg, Germany.
b) Author to whom correspondence should be addressed. Electronic mail: tobias.heindel@tu-berlin.de

In recent years, significant advancements have been achieved in the field of solid-state based single-photon emitters,[1] which enabled e.g. the demonstration of boson sampling[2,3] or heralded spin-spin entanglement[4] using semiconductor quantum dots (QDs). Such experiments crucially depend on efficient sources of indistinguishable photons. By integrating single semiconductor QDs within microcavities, quantum light sources with high extraction efficiencies, close to zero multi-photon emission probability and a high degree of photon indistinguishability can be realized.[5-7] For advanced quantum information processing schemes, however, a crucial resource is the indistinguishability of photons originating from remote emitters. So far, only few experiments reported two-photon interference (TPI) from spatially separated QD-based single-photon sources.[8-12,4,13] Deterministic sources have only been applied very recently,[14] exploiting high-Q microcavities. The corresponding narrow-band enhancement, however, makes them fragile and prevents applying them for schemes requiring efficient photon extraction of more than one QD state in advanced quantum communication schemes based on entanglement distribution. Waveguide approaches, such as photonic nanowires,[15] offer large photon extraction efficiencies in a broad spectral range. For this type of structures, however, high degrees of photon-indistinguishability still need to be demonstrated. Using geometrical approaches, such as microlenses, one can simultaneously achieve significantly enhanced photon extraction efficiency (compared to bulk material), large spectral bandwidth as well as high single-photon purity in terms of the photon-indistinguishability.[16] In addition, by applying an in-situ electron beam lithography approach the QD microlens' emission energy can be chosen deterministically.[17] This represents a highly desirable requirement for efficient scaling of photonic quantum information networks.

In this work, we demonstrate TPI from two spatially separated, deterministic single-photon sources based on QDs embedded within microlens structures. Exploiting in-situ electron-beam lithography in combination with cathodoluminescence spectroscopy enables us to integrate QDs with identical target wavelength deterministically within monolithic microlenses. By applying a quasi-resonant excitation scheme, we are able to demonstrate a TPI visibility of 49% and 22% for consecutive emitted photons of single emitters. The quantum interference of remote sources yields a TPI visibility of $V_{\text{remote}} = 29\%$ and proves our approach to be attractive for the realization of entanglement swapping based on entangled photon pairs employing the biexciton-exciton radiative cascade.

The samples used in this work were grown by metal-organic chemical vapor deposition (MOCVD) on GaAs (001) substrate. The structure consists of a lower distributed Bragg reflector (DBR) with 23 pairs of AlGaAs/GaAs located 65 nm beneath the InGaAs QD layer, which is capped with 400 nm GaAs. QD microlenses are fabricated via 3D in-situ electron-beam lithography combined with cathodoluminescence spectroscopy.[16] Here, microlenses are patterned at the positions of selected QDs with a spectral accuracy of about 0.4 nm[18] by locally inverting the electron beam resist. During fabrication, the electron doses for mapping and 3D patterning are carefully chosen to ensure a high process yield.[19] Using the microlens fabrication process described above, we achieved photon extraction efficiencies up to 29%.[20] Experiments were performed in a confocal microphotoluminescence (µPL) setup with the samples mounted onto the cold finger of liquid helium flow cryostats. For the selection of a suitable pair of QD microlenses from both samples, labeled QD 1 and QD 2, we considered the matching of the energetic splitting between *s*- and *p*-shell. This assures that both QDs can be efficiently excited by pumping them into an excited state at 909.5 nm using a mode-locked Ti:sapphire laser operating in picosecond-mode at a repetition rate of 80 MHz. The spectrally filtered QD emission can be coupled into fiber-based Hanbury-Brown and Twiss (HBT) or Hong-Ou-Mandel (HOM)-type setups with a timing resolution of 350 ps. For details on the HBT- and HOM-type measurements on the individual QDs, we refer to Ref. 16. For the actual interference experiments of remote emitters, both QD microlenses were operated in two cryostats separated by a distance of 1.5 m (see schematic in Fig. 1(a)). The photon streams emitted by both sources are orthogonally polarized (using two λ/2-waveplates), spatially superimposed at a polarizing beam splitter (PBS), energetically filtered in a grating spectrometer and coupled into a Mach-Zehnder-Interferometer (MZI) made out of polarization-maintaining single-mode fiber. The first beam splitter in the MZI is a PBS, and hence spatially splits the orthogonally polarized photons of QD 1 and QD 2 into both interferometer arms. Here, one arm of the MZI includes a fiber-based variable optical delay line for a precise control of the relative temporal delay, while the other arm contains a λ/2-waveplate for a controlled switching of the polarization from cross- to co-parallel polarization configuration. Next, the photons from QD 1 and QD 2 impinge on a non-polarizing beam splitter (BS) and are finally detected at the two output ports using silicon-based single-photon counting modules (SPCM). Additionally, a filter wheel introduced within one beamline allows for a careful balancing of the intensities of both photon streams at the second beam splitter inside the MZI, which is crucial for the observation of the TPI effect.

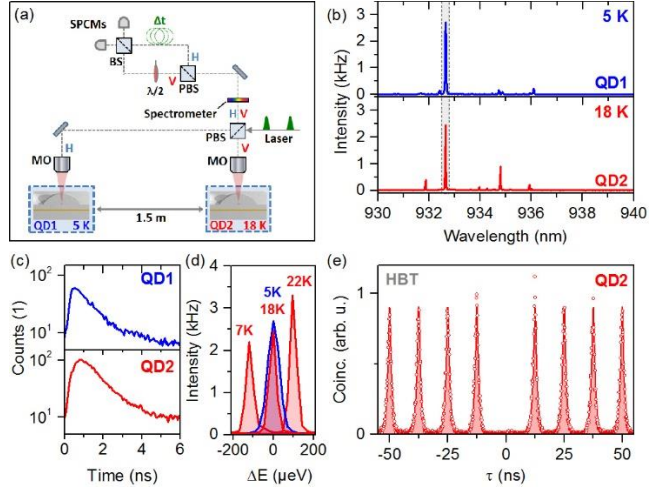

FIG. 1. (a) Setup used to observe TPI from remote microlenses. A single excitation laser pumps the QDs simultaneously in their p-shell. The polarizing beam splitters (PBS) combine (split) the photon streams before (after) the spectrometer. Inside the Mach-Zehnder-Interferometer (MZI) a half-wave-plate is used to change between co- and cross-polarized photons. An optical delay allows for temporal matching of the photons at the last non-polarizing beam-splitter (BS), where TPI takes place. (b) µPL spectra of QD 1 and QD 2 with emission from identical wavelengths. (c) Lifetime measurement for QD 1's and QD 2's excitonic state at 1.3292 eV emission energy. (d) By tuning the temperature of QD 2, its energy can be fine-tuned to match QD 1. (e) Measured HBT autocorrelation function for QD 2 showing excellent multi-photon-suppression.

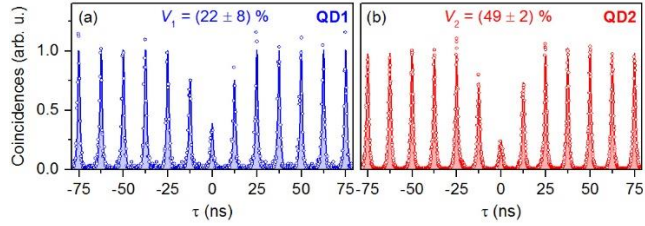

FIG. 2. Measured two-photon interference coincidence histograms $g^{(2)}_{HOM}(\tau)$ and model fit for (a) QD 1 and (b) QD 2 at $T = 10$ K. Suppression of the central peak indicates emission of indistinguishable photons. Solid lines represent a model fit considering the timing resolution of the detection system.

Fig. 1(b) shows the μPL emission spectra of both microlenses QD 1 and QD 2. Varying the sample temperature, the emission energy of QD 2 was fine-tuned to match the energetic position of QD 1. This fine-tuning is depicted in Fig. 1(d) for three different temperatures. Matching is obtained at temperatures of $T_1 = 5$ K and $T_2 = 18$ K. We determined the radiative lifetimes of both excitonic complexes via time-resolved μPL measurements (see Fig. 1(c)) to be $\tau_1 = (0.98 \pm 0.02)$ ns and $\tau_2 = (1.03 \pm 0.03)$ ns. Measuring the autocorrelation function $g^{(2)}_{HBT}(\tau)$ of microlenses in the HBT-type setup, excellent single-photon emission with an uncorrected antibunching of $g^{(2)}_{HBT}(0) = 0.01 \pm 0.01$ is revealed, as displayed exemplarily for QD 2 in Fig. 1(e).

Next, we determined the photon-indistinguishability of each individual QD microlens using the HOM-type setup with the photons being emitted with a temporal separation of 12.5 ns. As can be seen from the coincidence histograms for parallel polarized photons in Fig. 2(a) and (b), the coincidences at zero time delay are strongly reduced, proofing quantum mechanical interference of consecutively emitted photons. To extract the $g^{(2)}_{HOM}(0)$ values from the coincidence histogram we fitted the raw measurement data using the model function from Ref. 21 accounting for the setup's timing resolution. The corresponding visibilities can then be calculated by $V = 1 - 2 \times g^{(2)}_{HOM}(0)$ assuming $g^{(2)}_{\perp}(0) = 0.5$ in the case of cross-polarized photons. From the data in Fig. 2 we obtain values of $V_1 = (22 \pm 8)\%$ and $V_2 = (49 \pm 2)\%$, for QD 1 and QD 2 respectively at $T = 10$ K. The moderate visibilities observed for both microlenses can mainly be attributed to stochastic charge noise in the QD's vicinity – an effect which is significantly reduced for shorter photon emission time intervals.[22]

In the following, we experimentally address the mutual interference of photons emitted by the spatially separated QD-microlenses. For this purpose, we interfere the photons emitted by both QD microlenses at the second beamsplitter of the MZI. Fig. 3 presents the coincidence histogram $g^{(2)}_{HOM}(\tau)$ from which a uncorrected TPI visibility of $V_{1+2} = (29 \pm 6)\%$ is extracted by fitting. The visibility observed for our remote sources is therefore already competitive with the state-of-the-art reported for deterministic devices without post-selection ($(40 \pm 4)\%$).[14] Applying resonant excitation schemes, we anticipate significantly improved performance on the level demonstrated for non-deterministic devices ($(91 \pm 6)\%$).[4] The theoretically expected TPI visibility $V^{theo}_{1+2}$ for zero spectral detuning of both emitters can be deduced from the radiative lifetimes $\tau_i$ and pure dephasing rates $\tau_i^*$ of the individual emitters according to:[14]

$$V^{theo}_{1+2} = 4 \times \frac{1}{\tau_1 + \tau_2} \times \frac{1}{\left(\frac{1}{\tau_1} + \frac{1}{\tau_2} + \frac{2}{\tau_1^*} + \frac{2}{\tau_2^*}\right)} \quad (1)$$

Here, we derive the dephasing rates $\tau_i^*$ by extrapolating the individual TPI visibilities $V_i$ measured at $T = 10$ K (cf. Fig. 2) for the temperatures present in the actual measurement of $V_{1+2}$ (QD 1 at 5 K, QD 2 at 18 K). According to the temperature dependence reported in Ref. 22, we assume $\tilde{V}_1 \approx V_1 = (22 \pm 8)\%$ for QD 1, while we estimate a visibility of $\tilde{V}_2 \approx (26 \pm 10)\%$ for QD 2, accounting for the increased contributions of phonon induced pure dephasing at $T = 18$ K. The pure dephasing times $\tau_i^*$ are then derived from the individual TPI visibilities $\tilde{V}_1$ and $\tilde{V}_2$ as well as the radiative lifetimes $\tau_1$ and $\tau_2$ (extracted from Fig. 1(c)):

$$\tilde{V}_i = \frac{\tau_i^*}{\tau_i^* + 2\tau_i} \quad (2)$$

This yields pure dephasing times of $\tau_1^* = (0.55 \pm 0.14)$ ns and $\tau_2^* = (0.73 \pm 0.38)$ ns. Here, we would like to point out that the TPI visibility $V^{theo}_{1+2}$ from remote QDs arises from two completely uncorrelated semiconductor environments. Hence, one has to consider pure dephasing times $\tau_1^*$ and $\tau_2^*$ from measurements not affected by noise correlations, such as spectral diffusion. To rule out such correlations, we intentionally measured the visibilities of the individual emitters at a large pulse separation (12.5 ns).[22] Using the derived values for $\tau_i$, and $\tau_i^*$ with Eq. 1, we obtain $V^{theo}_{1+2} = (24 \pm 13)\%$, in agreement with our experimentally determined value $V_{1+2}$.

In summary, we experimentally demonstrated TPI from remote deterministic single-photon sources. For this purpose, we fabricated monolithic QD microlenses emitting at identical target wavelengths, employing 3D in-situ electron beam lithography including pre-selection of suitable QDs. These microlenses show single photon emission with $g^{(2)}_{HBT}(0) = 0.01$ and using two spatially separated sources an uncorrected TPI visibility of $V_{1+2} = 29\%$ is measured. Recently we have shown that QD microlens are compatible with a resonant two-photon-excitation scheme.[23,24] Moreover, QD microlenses can be combined with strain-tuning techniques, which could enable spectral fine-tuning of the emitter without affecting the quantum optical properties. Even more advanced is a recent technique showing the independent control of emission wavelength of two QDs on the same sample structure.[25] Both approaches promise to yield an even higher TPI visibility in future experiments. As QD microlenses increase the QDs' extraction efficiency over a broad spectral range, they are an attractive platform to realize efficient entangled photon pair sources employing the naturally entangled photons from the biexciton-exciton cascade. Having shown that TPI from remote QD microlenses is possible we envisage entanglement swapping between two such entangled photon pair sources.

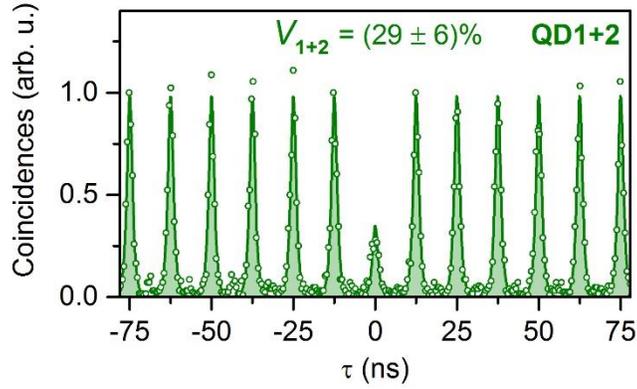

FIG. 3. Measured coincidence histogram $g^{(2)}_{HOM}(\tau)$ resulting from the mutual interference of the two remote QD single-photon sources. The solid line represents a fit to the raw measurement data. We observe a photon-indistinguishability with an uncorrected visibility of $V_{1+2} = 29\%$.

## ACKNOWLEDGMENTS

The research leading to these results has received funding from the German Research Foundation via CRC 787 and from the European Research Council under the European Union's Seventh Framework ERC Grant Agreement No. 615613. We acknowledge expert sample processing by R. Schmidt.